\documentstyle[]{article}

\begin{document}
\setlength{\unitlength}{0.17cm}
\title{Quantum vs stochastic processes and the role of complex numbers}
\author{Charis Anastopoulos \thanks{anastop@phys.uu.nl}, \\
\\Spinoza Instituut, Leuvenlaan 4, \\
3584HE Utrecht, The Netherlands  } \maketitle

\renewcommand {\thesection}{\arabic{section}}
 \renewcommand {\theequation}{\thesection. \arabic{equation}}
\let \ssection = \section
\renewcommand{\section}{\setcounter{equation}{0} \ssection}

 {\em Invited
contribution to Peyresq VII conference.}
\\ \\
\begin{abstract}
We argue that the complex numbers are an irreducible object of quantum probability:
this can be seen in the measurements of geometric phases that have no classical
probabilistic analogue. Having the complex phases as primitive ingredient implies that
we need to accept non-additive probabilities. This has the desirable consequence of
removing constraints of standard theorems about the possibility of describing quantum
theory with commutative variables. Motivated by the formalism of consistent histories
and keeping an analogy with the theory of stochastic processes, we develop a
(statistical) theory of quantum processes: they are characterised by the introduction
of a ``density matrix'' on phase space paths (it thus includes phase information) and
fully reproduces quantum mechanical predictions. We can write quantum differential
equations (in analogy to Langevin equation) that could be interpreted as referring to
individual quantum systems. We describe the reconstruction theorem by which a quantum
process can yield the standard Hilbert space structure if the  Markov property is
imposed. We discuss the relevance of our results for the interpretation of quantum
theory (a sample space is possible if probabilities are non-additive) and quantum
gravity (the Hilbert space arises here after the consideration of a background causal
structure).
\end{abstract}
\section{Introduction}
{\em ... Is what the matrix-physicists and q-number-physicists say true - that the
 wave equation describes only the behavior of a statistical ensemble, just like the so-called
 Fokker differential equation?...} \\
\hspace{5cm} Schr\"odinger to Planck, 1927
\\ \\
This quotation is from a letter Schr\"odinger sent to Planck \cite{Schr} after  the
5th Solvay Congress in 1926. In this congress the matrix mechanics of Heisenberg and
the wave mechanics of Schr\"odinger had faced each other and were reconciled by the
brilliant idea of Born to interpret Schr\"odinger's wave function as corresponding to
probability.

This proved to be one of the last pieces in the puzzle towards the development of a
full-fledged quantum theory. But Schr\"odinger was not satisfied with this
interpretation; he distrusted the
 philosophy underlying  the physical ideas of Heisenberg, Born
and Bohr (``the matrix-physicists and q-number-physicists'' in the quotation). In
fact, he wanted to interpret the wave function as a physical wave, but he could not:
the wave function was defined on the configuration space and not on the physical
space, as any physical wave ought to.

Fokker's equation, to which Schr\"odinger refers  is known today as  the Fokker-Planck
equation. It is the equation that describes the time evolution of the probability
density in Brownian motion. For Schr\"odinger it was natural to consider that a
probabilistic interpretation of the wave function would amount to its having a
physical function analogous to that of  a  classical probability distribution. In such
a  case one could say that the wave function provides the description of a statistical
ensemble rather than an individual quantum system.

But the interpretation of Schr\"odinger's equation as a Fokker-Planck  type of
equation did not gain ground and for good reason. The structure of these equations are
very distinct. The Fokker-Planck equation  reads
\begin{equation}
\frac{\partial}{\partial t} \rho = {\cal L} \rho,
\end{equation}
where $\rho$ is a probability distribution and ${\cal L}$ a linear operator  in the
space, where the $\rho $ live. Schr\"odinger's equation is of course
\begin{equation}
i \frac{\partial}{\partial t} \psi = H \psi
\end{equation}
$H$ is the Hamiltonian, a self-adjoint operator on the Hilbert space, where the $\psi$
live. The difference between these two equations and what distinguishes between
classical and quantum probability is the presence of the complex unity $i$. Well, you
might argue that complex numbers are not really measurable quantities, just a
convenient device to simplify the writing of equations; indeed, Schr\"odinger's
equation can be written without the use of $i$ by splitting it into two equations, one
for the real and one for the imaginary part of $\psi$. However, physical observables
arise out  of the probability distribution $\rho = |\psi|^2$, so even if we forego the
use of complex numbers they will reappear in the form of the $U(1)$ invariance of the
probabilities.

In effect, what distinguishes quantum  from classical probability  is the presence of
the complex numbers, not so much in the dynamics, but as the $U(1)$ symmetry of the
probability assignment.

Coming back to the Fokker-Planck equation, we note that they  it is a part  of the
more general theory of stochastic processes. The stochastic processes can describe the
dynamics in two ways: either through the Fokker-Planck equation by writing a
deterministic equation for the probability density that refers to an ensemble of
physical systems, or through stochastic differential equations (like the Langevin
equation) that can be interpreted as referring to the behaviour of an {\em individual}
system evolving under some random forces. In classical probability, we then have the
following diagram
\\ \\
{\large Fokker-Planck eqn.   $ \leftarrow $  {\em stochastic processes } $\rightarrow$
Langevin eqn.}
\\ \\
If we try  to interpret Schr\"odinger's equation as analogous to the Fokker-Planck
equation, we immediately see that we lack both the general theory in which it will be
embedded and the description (analogous to differential equations) that could be
interpreted as referring to individual quantum systems. So the corresponding diagram
has two gaps in it
\\ \\
{\large  Schr\"odinger's eqn. $ \leftarrow  X_1$ theory    $\rightarrow X_2$  eqn.}
\\ \\
 Nelson attempted to fill the gap  in his stochastic mechanics
programme \cite{Nel85, GORW78}, where for $X_1$ he considered again the theory of
stochastic processes and for $X_2$ again stochastic differential equations
\footnote{He actually considered the Madelung equations that are  derived from
Schr\"odinger's equation and the probability rule.}. However, this description cannot
account for all quantum phenomena as it violates Bell's theorem. In fact, no
stochastic process can reproduce all quantum mechanical predictions \cite{GHT79}.

In this talk we are going to show how to fill in the previous diagram, without
diverging from the predictions of standard quantum theory; we are going to write a
class of theories which is modelled on stochastic processes but are not stochastic
processes themselves, since they intrinsically incorporate the appearance of the
complex numbers (or if you prefer a $U(1)$ symmetry). This class of theories we shall
call {\em quantum processes}. They have been developed in a chain of argumentation
starting from the consistent histories approach to quantum theory and the presence of
the geometric phase. Quantum processes is the $X_1$ in our previous diagram; they can
be unravelled to write full analogues of the stochastic differential equations, which
could perhaps be interpreted as referring to individual system.

In general, our approach is very close to the ideas about quantum theory,  that
 Einstein expressed in his later years \cite{Ein} (in which it was made clear that his
 disagreement with quantum theory was not on reasons of determinism, but on reasons of
 realism)
 \\ \\
``... {\em ... The concept that the $\psi$-function completely describes the physical
behaviour of the individual single system is untenable. (...) But if one regards the
$\psi$-function as the description of an ensemble, it furnishes statements that
correspond satisfactorily to those of classical mechanics and at the same time account
for the quantum structure of reality.... }...''
\\ \\
In this paper we demonstrate, that this attitude is not forbidden  by any non-go
theorem of quantum mechanics; indeed at the mathematical level we can even write
equations that could be interpreted as referring to individual systems. And, at least
for the author, this attitude is in no way incompatible with the basic physical
insights, that were put into the structure of quantum theory by Heisenberg and Bohr.

\section{Classical vs quantum probability}

Quantum theory was built in the late twenties and by the  early thirties its basic
principles and structures had been put in place. During the same period, classical
probability was set in a solid axiomatic framework in the work of Kolmogorov.
Kolmogorov founded probabilistic calculus on measure theory and thus removed
ambiguities plaguing other  intuitive approaches to probability. In particular,
measure theory provided  a framework, by which the theory of stochastic processes
could be rigorously developed.

This was arguably one of the most important results in 20th century's  applied
mathematics; in particular, it influenced the more mathematically-minded people that
worked on the foundations of quantum theory. In particular, it was soon apparent that
probability theory and quantum mechanics seemed to share the same basic ``physical''
concepts, even if their mathematical implementation is distinct. Classical probability
theory is defined on a sample space $\Omega$, which is an ordinary set (often a
manifold), while quantum theory is defined on a {\em complex} Hilbert space $H$.

A theory such as classical probability is a mathematical framework
 by which physical phenomena can be modeled, so that the statistical results
 (mean values, ratios of events)   measurements
  can be predicted. So is quantum theory, at least in the Kopenhagen interpretation.
  Both theories employ the notions of \\ \\
1. {\bf Observables:} An observable is a physical quantity, whose value we measure.
It is assumed that what we observe can only be a real number, since it is with respect
to real numbers (distances in dials) that we encode all experimental information. It
has to be a single-valued object, so that when we have a concrete measurement
situation there will be no ambiguity as to the quantity we measure. Classically an
observable is a measurable function on $\Omega$, while quantum mechanically a
self-adjoint operator on $H$.
\\ \\
2. {\bf Events:} As events we characterise the possible outcomes of individual
experiments: an event corresponds to a property that is verified by an experiment,
e.g. the particle passed through a slit located in that particular place. Quantum
mechanical events are represented by projection operators and  classical ones  by
measurable subsets of the sample space.
\\ \\
3. {\bf States:} A state corresponds to the preparation of the physical  system,
before the measurement is carried out. As such, it contains information as much about
the nature of the physical system as about the preparation procedure that was used for
the experiment. A state should be a mathematical object that would be able to provide
all information that can be accessed experimentally. If we think that in experiments
we can determine mean values for observables and (perhaps) probabilities to events,
the state should be an object that provides these.  In classical probability theory a
state corresponds to  a probability distribution, while in quantum theory (thanks to
Gleason's theorem) by a density matrix. Note, however, that in representing the
operational notion of the state by such mathematical objects we assumed that the
probabilities that correspond to physical systems satisfy a number of properties. Most
important of them is the additivity condition, that if $A$ and $B$ are two independent
events then $p(A \cup B) = p(A) + p(B)$. This is a theoretical ``prejudice'' rather
than an axiom that arises naturally out of the consideration of the measurement
processes.
 \\ \\
In any case, starting from the work of von Neumann, Birkhoff, Wigner  and Jordan
\cite{vNWJ,BvN}   in the  thirties,   an ever increasing number of mathematical
physicists considered quantum theory to be nothing but a generalised probability
theory, i.e. a theory sharing the same basic objects as classical probability theory.
This attitude is apparent in many schemes of axiomatisation of quantum theory, that
want to get rid of the (mathematically)  unintuitive  structure of a Hilbert space
(while keeping the non-commutativity of observables), by substituting it with
something simpler: algebras of observables in the $C^*$-algebraic approach, lattice of
propositions in quantum logic(s), convex state spaces in the operational approach.

However, we still feel the need to ask the question, whether {\em
quantum theory is nothing but a generalised probability  theory.}\\ \\
Our answer here is that {\bf it is not.}
\\ \\
The analogy between classical probability and quantum mechanics stops, when we
consider properties of the quantum system at more than one moment of time. Because in
this case
\\ \\
1. Interference phases appear that have no analogue in classical probability theory.
They are closely related to the geometric
phases. Most important of all, they are measured as a statistical object.\\
2. The probabilities for properties defined on more than one
moment of time are {\em non-additive}. \\
3. The interference phases and the non-additive probabilities are
closely related. \\
4. The natural correlation functions of the observables are probabilities generically
complex-valued.
\\ \\
It seems that in the case of studying properties at more than one moment of time, the
complex numbers inherent in the structure of the quantum mechanical Hilbert space
become manifest.

\subsection{Non-additive probabilities}
To explain the non-additive probabilities in quantum theory let us consider the
following type of experiment. We have a source $S$, which emits some particles (our
individual quantum systems) prepared in a well-defined state $| \psi \rangle$. The
ensemble of systems is then represented by a beam; we let this beam cross two filters
represented by $P$ and $Q$. A filter is an object that lets particles of the beam pass
if they satisfy a certain property or, in other words, if a particular event occurs.
In quantum theory a filter is represented by a projection operator, hence $P$ and $Q$
are projectors.

Now, after the second filter $Q$ the beam falls into a detector $D$ and we can measure
the particles that have crossed the beams. Note that in the diagram the distance
between the source, the filters and the detector is assumed to represent time
intervals.

 If the source emitted $N$ particles and the detector detected $n$, then for large $N$ the ratio $n/N$ ought to converge to a given number, which would be the probability for the particles prepared in the state $| \psi \rangle$  to pass through the two filters.

The rules of quantum theory give that this probability must be equal to
\begin{eqnarray}
 p(P,t_1;Q,t_2)&=& \langle \psi|P(t_1)Q(t_2) P(t_1)| \psi \rangle
\\ \nonumber P(t) : &=& e^{iHt}P e^{-iHt} ,
\end{eqnarray}
where $H$ is the Hamiltonian that describes the self-dynamics of the individual
system.

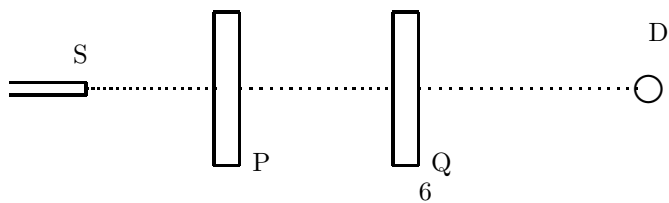
\begin{picture}(16,16)
\thicklines
 \put (0,-0.5) {\line(1,0){6}}
 \put (0,0.5) {\line(1,0){6}}
\put (6,-0.5) {\line(0,1){1}} \put (16, -6) {\line(0,1){12}} \put (18,-6)
{\line(0,1){12}} \put (16,-6) {\line(1,0){2}} \put (16,6) {\line(1,0){2}} \put (30,
-6){\line(0,1){12}} \put (32,-6) {\line(0,1){12}} \put (30,-6){\line(1,0){2}} \put
(30,6) {\line(1,0){2}} \put (50,0){\circle{2}} \qbezier[18] (6,0) (10,0) (16,0)
\qbezier[17] (18,0) (25,0) (30,0) \qbezier[22] (32,0) (41,0) (49,0)
 \put(5,2){S}
\put(19,-6.4){P} \put(33,-6.4){Q} \put(50,3.7){D}
\end{picture}
\\ \\ \\ \\ \\
Now consider the following three experiments. In experiment number 1, we put in as a
first filter  $P_1$ and as second one  $Q$. In experiment number 2, we put as a first
filter $P_2 = 1 - P_1$, i.e. the filter corresponding to the property complementary of
that of $P_1$ and keep $Q$ as a second filter . In experiment number 3, we simply have
filter $Q$. If we use the rule (2.1) for the measured probabilities we see that
\begin{equation}
p_3 \neq p_1 + p_2.
\end{equation}
This implies that the quantum mechanical probabilities do not satisfy the additivity
condition. This simple experiment is a manifestation of the more general fact for
quantum theory: {\em  experiments involving properties  of the system at more than one
moment of time cannot, in general,  be modeled by an additive probability measure}. So
quantum probability for histories does not satisfy the basic Kolmogorov axioms for
probabilities.

\subsection{Consistent histories}

Well, why would we mind if the probabilities are non-additive? Non-additivity of
implies you lose an important rule of inference. Consider that you have two exclusive
events $A$ and $B$, that are also exhaustive (one or the other can happen and nothing
else). If we measure the probability for $A$ and we find $p(A) = 1$, then we can be
certain that the event $B$ would never occur in any repetition of this experiment,
since $p(B)$ has to equal zero. If probabilities are non-additive, then the fact that
$p(A) = 1$ does not imply that $p(B) = 0$ and the event  $B$ would take place in the
experiments with non-zero frequency.

In a nutshell, even if the probability of an event is $1$, we cannot preclude that its
complement will never happen. Is this so bad? Not really if we have an operational
stance, that quantum theory describes experiments in ensembles. However, it could be
problematic if one wants to claim that the present formalism of quantum theory
provides a theory for individual closed systems, because it would dramatically  limit
its predictability.

One way to resolve this problem (assuming it is a problem) is the consistent histories
approach. This was developed by Griffiths \cite{Gri84}, Omn\'es \cite{Omn8894},
Gell-Mann and Hartle \cite{GeHa9093, Har93a}. This work is motivated from its  elegant
formulation that has been developed by Isham \cite{I94, IL94, IL95} and Savvidou
\cite{Sav99a, Sav02b}.

As far as this issue goes, the key idea of the consistent histories approach is that
one can have additive probabilities if one is restricted within particular sets of
histories, known as {\em consistent sets}.  More precisely
\\ \\
 1. A general history $\alpha$ is represented by a collection of projection operators $\alpha_{t_i}$ at successive instants of time
\begin{equation}
\alpha = (\alpha_{t_1}, \alpha_{t_2} , \ldots, \alpha_{t_n}), \nonumber.
\end{equation}
2. From these operators we can  define an operator $C_{\alpha} = \alpha_{t_n}(t_n)
\ldots \alpha_{t_2}(t_2) \alpha_{t_1}(t_1)$ and a complex-valued  functional on pairs
of histories
 \begin{equation}
d(\alpha, \beta) = Tr(C_{\alpha} \rho_0 C_{\beta}^{\dagger}).
\end{equation}
This is known as the decoherence functional. \\ \\
3. If in a exhaustive and exclusive set of histories
\begin{equation}
d(\alpha, \beta) =0 , \hspace{3cm} \alpha \neq \beta
\end{equation}
then $d(\alpha, \alpha)$ is a probability for the history $\alpha$ and corresponds to
an additive probability measure. The satisfaction of condition (2.5) renders a set of
histories consistent.
\\  \\
The additivity of probabilities in each consistent set allows us to use the fact  that
probability one for an event means probability zero for its complement. This can be
exploited to define the logical/physical notion for implication. The idea is that if
we get by measurement a definite result for an {\em individual} system, we can employ
the definability of this implication to identify other properties that this system has
satisfied. However, implication is defined only within given consistent sets; hence
when we employ implication in different consistent sets we can derive contrary results
from the same definite (measured) event. This is a more general problem of realist
interpretational schemes \footnote{In our use, we employ the word realist to denote
the attitude that the quantum mechanical formalism refers to properties of individual
systems.} and is related to the Kochen-Specker theorem.

However, the {\em formalism} of consistent histories makes sense irrespective of the
interpretation: it can equally well be considered in a Copenhagen framework, in which
we are  content to provide predictions for outcomes of ensemble measurements. The
merit of the formalism lies in the fact that it allows the description of quantum
systems using {\em covariant objects} (histories) \cite{Har93a}. As such it seems more
adequate to deal with basic problems in the quantisation of the gravitational field,
such as the problem of time (see \cite{Sav01} for this perspective). Moreover, the
object that was introduced by the consistent histories approach, the decoherence
functional  is very convenient: it will be shown to contain the full information that
can be extracted from a quantum process, even its non-probabilistic aspects.

\subsection{Complex-valued correlation functions}

Whenever we have a probabilistic system, there is a well defined prescription by which
the {\em temporal correlation functions} can be determined. Assume we have an
observable $A$, that can be spectrally analysed as $A = \sum_i a_i P_i$, where $P_i$
are filters that can be experimentally employed. Now we repeat the experiment in
figure 1 with $P_i$ for $P$ and $P_j$ for $Q$  for all possible combinations of $i$
and $j$. We can then measure the probabilities $p(i,t_1;j,t_2)$.

The statistical correlation function is then
\begin{equation}
\langle A_{t_1} A_{t_2} \rangle_S = \sum_{ij} a_i a_jp(i,t_1;j,t_2).
\end{equation}
It is clearly a real number. However, it is an object that cannot be naturally written
in terms of the operators $A$. Moreover, it depends sensitively in the resolution of
the observable in terms of the spectral projections (unlike statistical correlation
functions in classical probability theory).

From quantum theory the natural object for the correlation function (that does not
depend on the spectral resolution of the observable) is
\begin{equation}
\langle A_{t_1} A_{t_2} \rangle_Q = \langle \psi(t_1) | A e^{iH(t_2 - t_1)} A
e^{-iH(t_2 - t_1)} | \psi(t_1) \rangle,
\end{equation}
or the corresponding time-ordered function.

This is generically a complex-valued object and has no natural operational
interpretation as the statistical correlation function.

We should remark that temporal correlation functions have actually been measured in
quantum optics (see for instance \cite{WM}). In that case the relevant observable is
the photon number; this, however, commutes with the electromagnetic field's
Hamiltonian. This implies that the statistical and quantum correlation functions
coincide.

So one may pose the question, {\em why does quantum theory give as natural correlation
functions, ones that are not operationally implementable?} This is not a very sharp
question, as the answer might be simply that there is no {\em a priori} reason to
expect that it would be otherwise. However, we are going to show that there is a
deeper reason and this is that quantum correlation functions have information from
measurable quantities that do not correspond to probabilities: {\em interference
phases}.

\section{Interference phases}
The point is that quantum theory predicts other physical quantities that  can be
determined statistically, but are not probabilities. These are the quantum phases and
more precisely the geometric phases, paradigmatic example of which are  the
Bohm-Aharonov \cite{BoAh59} and the Berry phase \cite{Ber84}.

Let us recall the measurement of the Bohm-Aharonov phase
\\ \\
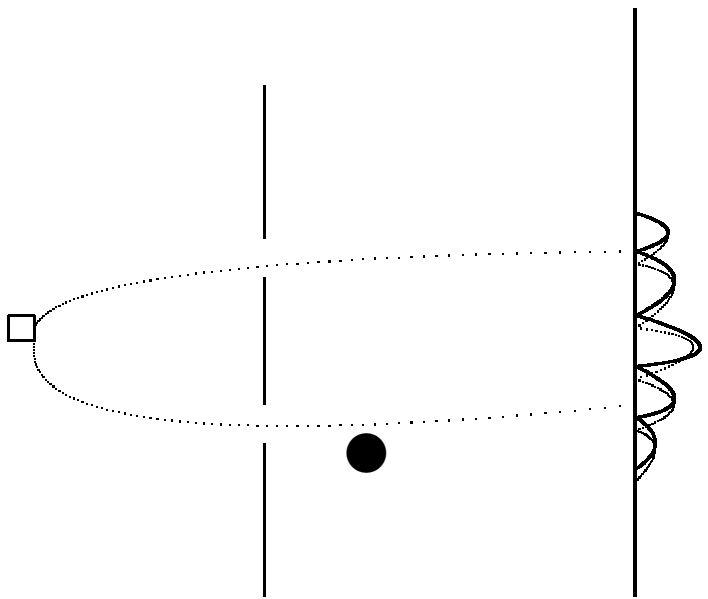
\begin{picture}(15,15)
\thicklines \put (0,-20) {\line(0,1){2}} \put (0,-18) {\line(1,0){2}} \put (2,-18)
{\line(0,-1){2}} \put (2,-20) {\line(-1,0){2}} \put (20,0) {\line(0,-1){12}} \put
(20,-15){\line(0,-1){10}} \put(20,-28) {\line(0,-1){12}} \put(55,6) {\line(0,-1){46}}
\put(49,6) {\line(0,-1){46}} \put(28,-28.8) {\circle*{3}} \thinlines
\bezier{80}(2,-20)(0,-13.5)(49,-13) \bezier{80}(2,-20)(0,-30)(49,-25) \thicklines
\qbezier(49,-10)(54,-11.5)(49,-13) \qbezier(49,-13)(55,-15)(49,-18)
\qbezier(49,-18)(59, -21)(49, -22) \qbezier(49,-22)(55, -25)(49, -26)
\qbezier(49,-26)(52,-28)(49,-30) \thinlines \bezier{40}(49,-10)(54,-11.5)(49,-14)
\bezier{40}(49,-14)(55,-15)(49,-19) \bezier{40}(49,-19)(58, -20)(49, -23)
\bezier{40}(49,-23)(55, -25)(49, -27) \bezier{40}(49,-27)(52,-28)(49,-31)
\end{picture}
\\ \\ \\ \\ \\ \\ \\ \\ \\ \\ \\ \\ \\ \\ \\ \\ \\ \\
The basic configuration is that of a two-slit experiment. We let a prepared  state of
electrons cross through two slits and then measure the interference pattern on a
screen. Having stored that in memory, we repeat the experiment by putting a solenoid
(with some magnetic flux) behind one of the slits (such that the beam does not cross
it). We observe a shift into the interference pattern, which is essentially
proportional to the Bohm-Aharonov phase induced by the magnetic flux.

Two remarks must be made for this experiment, that are valid for all  experiments
measuring geometric phases. First, that the Bohm-Aharonov phase is a {\em statistical}
object: it is measured in terms of an interference pattern, which is present only when
a large number of electrons (thought of as corresponding to a statistical ensemble)
are left to interfere. If we carried out the experiment with a single electron, there
would be nothing to measure.

The second remark is that the geometric phase can be determined from the  study of the
interference of  two  beams  with different history. This is reminiscent of the
decoherence functional, which assigns a complex-valued object to a pair of histories.
This suggests that  the decoherence functional is somehow related to the geometric
phases. This loose connection will turn out to be  very precise: we can show that the
decoherence functional is actually constructed from the mathematical object that is
responsible for the presence of geometric phases: the Berry connection \footnote{This
is a $U(1)$ connection on the fiber bundle with base space the projective Hilbert
space and total space the Hilbert space of a quantum system \cite{Sim83, AnaAh87}.}
\cite{AnSav02}.

More strongly, we can explicitly describe, how the off-diagonal elements  of the
decoherence functional can be explicitly measured; the corresponding measurement
scheme is identical to the ones used for the determination of another version of the
geometric phase, known as the {\em Pancharatnam} phase \cite{Pan56, SaBha88}. This
phase essentially corresponds to the argument of the inner product between two states
$\phi \rangle$ and $|\psi \rangle$. It is manifested in the following generic
situations
\\ \\
1. Prepare two systems in the states $| \psi \rangle $ and $| \phi \rangle $. \\ \\
2. Perform on the beam corresponding to $| \psi \rangle$ the operation  $|\psi \rangle
\rightarrow e^{i \chi} | \psi \rangle$ for a controlled value of $\chi$.
\\ \\
3. Interfere the two beams to construct the beam
 $ |f \rangle = | \phi \rangle + e^{i \chi} | \psi \rangle$ and measure
 its intensity $ I = \langle f| f \rangle$. \\ \\
4. Repeat the experiment for the range of all values of $\chi$ and construct  the
function $I(\chi)$ giving the intensity of the measured beam as a function of
 the external parameter $\chi$. This equals
 \begin{equation}
 I(\chi) =  2 |\langle \psi| \phi \rangle| \cos ( \chi - \arg
 \langle \psi| \phi \rangle)
\end{equation}
 \\ \\
5. $I(\chi)$ takes its maximum value for $\chi = \arg \langle \psi| \phi \rangle$.
This value of $\chi$ is the Pancharatnam phase between the two beams.
\\ \\ \\
This procedure to measure the Pancharatnam phase has been performed  in neutron
interferometry \cite{WRFI97}. The difficult part is to perform step 2, i.e. to have a
controlled way to change the phase of an individual quantum state. This can be
achieved if  $| \psi \rangle$ is an eigenstate of a Hamiltonian (so that the phase
only depends on the number of periods the beam is left before interference).

Assuming that there is a prescription by which this  phase change can be performed,
 the above prescription can be used to measure the off-diagonal elements of the
decoherence functional.
\\ \\ \\
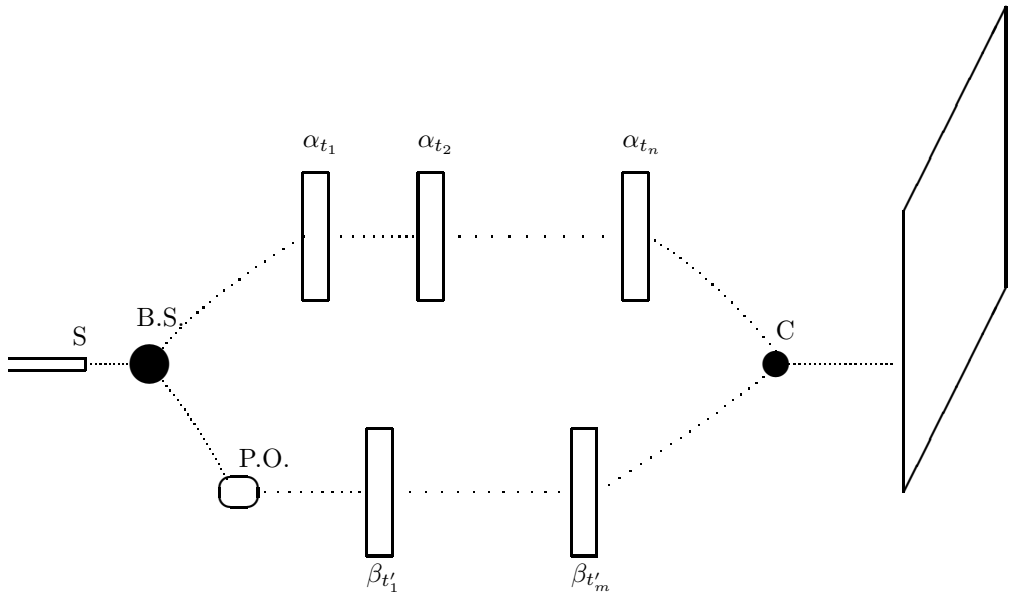
\begin{picture}(25,25)
\thicklines \put (0,-0.5) {\line(1,0){6}} \put (0,0.5) {\line(1,0){6}} \put (6,-0.5)
{\line(0,1){1}} \put (11, 0) {\circle*{3}} \put (23, 5) {\line(0,1){10}} \put (25,5)
{\line(0,1){10}} \put(23,5) {\line(1,0){2}} \put(23,15) {\line(1,0){2}} \put (32, 5)
{\line(0,1){10}} \put (34,5) {\line(0,1){10}} \put(32,5) {\line(1,0){2}} \put(32,15)
{\line(1,0){2}} \put (48, 5) {\line(0,1){10}} \put (50,5) {\line(0,1){10}} \put(48,5)
{\line(1,0){2}} \put(48,15) {\line(1,0){2}} \put(60,0) {\circle*{2}} \put
(70,12){\line(1,2){8}} \put(70,12){\line(0,-1){22}} \put(78,28){\line(0,-1){22}}
\put(70, -10){\line(1,2){8}} \put(18, -10){\oval(3,2.4)} \put (28, -15)
{\line(0,1){10}} \put (30,-15) {\line(0,1){10}} \put(28,-5) {\line(1,0){2}}
\put(28,-15) {\line(1,0){2}} \put (44, -15) {\line(0,1){10}} \put (46,-15)
{\line(0,1){10}} \put(44,-5) {\line(1,0){2}} \put(44,-15) {\line(1,0){2}} \thinlines
\qbezier[10] (6,0) (8,0) (9.5,0) \qbezier[20] (12,1.3) (15,5)  (23,10)
\qbezier[10](25,10) (30,10) (32, 10) \qbezier[10] (34,10) (40,10) (48,10) \qbezier[20]
(50,10) (55,7) (60,1) \qbezier[20] (12,-1.3) (15,-5)   (17,-9) \qbezier[10] (20, -10)
(25, -10)   (28,-10) \qbezier[12] (30, -10) (38, -10) (44, -10) \qbezier[18] (46, -10)
(54, -5)    (59, -1) \qbezier[20] (61,0) (65,0) (69,0) \put(5,1.5){S} \put (10,3)
{B.S.} \put (23,17) {$\alpha_{t_1}$ } \put (32, 17) {$\alpha_{t_2}$} \put (48, 17)
{$\alpha_{t_n}$} \put (60,2) {C} \put (18,-8) {P.O.} \put (28, -17) {$\beta_{t'_1}$}
\put (44,-17) {$ \beta_{t'_m}$}
\end{picture}
\\ \\ \\ \\ \\ \\ \\ \\ \\ \\ \\
Let us assume we have a source $S$ preparing particles in a state $| \psi \rangle$.
After exiting $S$, the beam enters a beam splitter B.S. One of its components then
enters a sequence of filters $\alpha_{t_1} \ldots \alpha_{t_n}$ and the other a
sequence of filters $\beta_{t'_1} \ldots \beta_{t'_m}$, before they are recombined at
C. The beam then propagates to a screen, where its intensity is measured. Now we
repeat this experiment many times, but at each time the second component of the split
beam has to pass through P.O. which performs the operation of phase change $ | \psi
\rangle \rightarrow e^{i \chi} | \psi \rangle$. Repeating the experiment for different
values of $\chi$, we  get a function $I(\chi)$, whose maximum determines a phase that
is the argument  of  the value of the decoherence functional between the histories
$(\alpha_{t_1} \ldots \alpha_{t_n})$ and $ (\beta_{t'_1}, \ldots, \beta_{t'_m})$.  The
modulus of the phase of the decoherence functional can easily be determined by the
maximum value of $I$ (see \cite{Ana02} for details).

In fact, it can be shown that  the decoherence functional contains all information
that can be obtained from experiments measuring either probabilities or relative
phases. The name decoherence functional is then, rather misleading. Gell-Mann and
Hartle introduced it as the object that provides information about  decoherence of
histories. However, its most important function is that it contains the information of
the relative phases, hence of coherence of histories. For this reason a more adequate
name for it would be the coherence functional or the phase functional.

It  can also be shown that the complex-valued temporal correlation functions  can in
principle be determined by the measurement of a sufficiently large number of
interference phases as  described above. In effect, if one can determine the
off-diagonal elements of a decoherence functional between a history $\alpha_{ij} =
(P_i, t_1; P_j, t_2)$ and the trivial history $\beta = 1$, then for the observable $A
= \sum_i a_i P_i$ the time-ordered two-point function will read
\begin{equation}
\langle A_{t_1}A_{t_2} \rangle_Q = \sum_{ij} a_i a_jd(\alpha_{ij},1),
\end{equation}
This gives an operational scheme for the determination of the quantum correlation
functions through measurements analogous to the one performed for the Pancharatnam
phase.
 The reader is referred to \cite{Ana01} for details.

\section{A framework for quantum processes}
\subsection{Phases as primitive ingredients}
Usually any discussion of the principles of quantum theory consider probabilities as
the basic objects that are predicted by the formalism. But, as we showed there is good
reason to consider (also) the phases as primitive ingredients of the formalism. We can
then attempt  to write an axiomatic scheme that achieves this.

In that case,  we must accept that the probabilities of the corresponding quantum
theory are non-additive, hence do not satisfy the Kolmogorov axioms. Accepting
non-additive probabilities has important consequences for the structure of the
resulting quantum theory. The theorems of Bell \cite{Bell64}
 and Kochen-Specker \cite{KoSp67} that forbid
hidden variable theories of reproducing the predictions of quantum theory assume that
the corresponding hidden variable theories are either deterministic or stochastic.
They do not forbid hidden variable theories that are modeled by a statistical theory
that is not described by Kolmogorov probability.

Hence, by accepting phases as primitive ingredients of our formalism, we might be able
to write a theory that reproduces the predictions of quantum mechanics, while having
observables like any classical theory, purely commutative objects. This is indeed
possible and we showed that in reference \cite{Ana01} by simply employing the Wigner
transform on the standard quantum mechanical decoherence functional. In other words,
the Hilbert space structure and the corresponding non-Boolean ( and non-distributive)
nature of the events (usually referred to as quantum logic) {\em is not necessary in a
theory that takes phases as primitive ingredients}.

Such a theory needs a mathematical structure analogous to that  of a decoherence
functional over a classical sample space. We shall explain in detail how such theories
are formalised and constructed: we shall call them theories of {\em quantum
processes}, because their mathematical structure is in many respects analogous to the
theory of stochastic processes.

 We will not attempt to write any interpretation of
quantum theory different from Copenhagen; we shall take a strictly operational stance
and simply consider that the theories of quantum processes refer only to measurement
situations and the probabilities and relative phases always make reference to
ensembles of individual quantum systems. We should, however, point out that our longer
term perspective is realist: we want to find a way to talk about the physics of
individual quantum system.  The present is not, however, a good moment for this
purpose.

\paragraph{Choosing the sample space}
If we want to write a theory that reproduces the results of quantum mechanics, while
having a classical sample space, we need to specify what this sample space would be.
One could take the stance that the proper quantum mechanical sample space consists of
variables very different from the ones that are naturally apparent to us: the true
degrees of freedom refer to a subquantum level of reality, quite removed from standard
physics (as for instance in \cite{thoof99}).

We shall take a more conservative approach here. He shall consider that the sample
space is essentially  the  phase space  of the corresponding classical system. There
are three reasons for our choice:
\\ \\
1. We can show that a theory of quantum processes on the  classical phase space allows
us to fully reproduce the predictions of standard quantum theory (this is an {\em a
posteriori} argument). \\ \\
2. Symplectic manifolds have a very rich geometric structure, that allows us to
reproduce many classic quantum mechanical results that seem to need {\em a priori} the
notion of the Hilbert space. Such is the case for Wigner's classification of particles
(from the geometric quantisation of  Konstant-Souriau \cite{Sour}) and many aspects of
the
spin-statistics relation \cite{Ana01pr}. \\ \\
3. One of our motivation for undertaking this line of research is the attempt to write
a quantum theory that has a pronounced spacetime character. In the closely related
scheme of histories quantisation, Savvidou showed that in the space of {\em phase
space histories}  one can always implement a symplectic action of the group of
spacetime diffeomorphisms \cite{Sav01} (in the case of general relativity it coexists
with the algebra of constraints obtained by the 3+1 decomposition). This is a very
important part for any quantum theory that wants to manifest the principle of general
covariance and holds only for phase space histories (not configuration space ones).
\\ \\ \\
These are the reasons that makes us consider the phase space histories as defining the
basic sample space of quantum theory. Our arguments are sufficient to establish the
naturality of our choice; of course, they are not necessary as they are directed by
choices related to our long term aims. In any case, the formalism we shall present
makes sense for any possible sample space.

\subsection{The basic axioms}
\subsubsection{Sample space, events, observables}
At the level of observables, the structure of our theory is identical with that  of
classical probability theory. That is, we assume the existence of a space $\Omega$ of
elementary alternatives. A point of $\Omega$ corresponds to the most precise
information one can extract from a measurement of the quantum system. Note, that at
this level we do not distinguish, whether $\Omega$ refers to properties of a systems
at one moment of time or to histories. Our definitions are general and only later
 shall we specify the history content.

 This space $\Omega$ has to be equipped with some
additional structure. In general, a measurement will yield some information stating
that the system was found in a given subset of $\Omega$. But not all subsets of
$\Omega$ are suitable to incorporate measurement outcomes.
 For instance, when we consider position it is physically meaningless to
 consider the subset of rational values of position (with respect to some unit).
 One, therefore needs to choose a family of subsets ${\cal C}$ of $\Omega$, that correspond to
 the coarse-grained information we can obtain about the physical systems. These sets are often
  called {\em events}. The   family ${\cal C}$ containing the events  has to satisfy some
   natural mathematical conditions \\ \\ \\
 i. $\Omega \in {\cal C}$: if an experiment is performed one of the outcomes will occur. \\ \\
ii. $\emptyset \in {\cal C}$: it is impossible that no outcome results if an
experiment
 is performed. \\ \\
iii. If $A \in {\cal C}$, then $\Omega - A \in {\cal C}$: if $A$ is a possible
measurement
 outcome then so can be its complement. \\ \\
iv. If $A, B \in {\cal C}$, then $A \cup B \in {\cal C}$ and $A \cap B \in {\cal C}$: unions and intersections of experimental outcomes are also possible  experimental outcomes. \\ \\
v. For countably many $A_n \in {\cal C}, n = 1, 2, \ldots$, $\cup_{n=1}^{\infty} \in
{\cal C}$. This is a technical condition particularly relevant when dealing with
 the case where $\Omega$ is a manifold.
  \\ \\
Equipping $\Omega$ with a choice of events  turns it into a {\em measurable space}.
\\ \\
 Since in experiments we eventually come to measure real numbers (or occasionally integers, which can be
embedded into the real numbers) the mathematical object that would represent the
notion of observable is a map from $\Omega$ to ${\bf R}$. However, not all possible
maps will do:
 the structures corresponding to measurable sets have to be preserved.
Such functions are called {\em measurable} and in the language  of probability theory
are known as  {\em random variables}. We shall denote the space they belong to in as
$F(\Omega)$.

Among all functions, important are characteristic functions of the various  subsets of
$\Omega$. These are defined as
\begin{eqnarray}
\chi_A(x) &=& 1 , x \in \Omega \\
 &=& 0 , x \notin \Omega
\end{eqnarray}

An important property of the characteristic functions is the following. If $\lambda$
is a possible value of a random variable $f$ and  $A_{\lambda} = f^{-1}({\lambda})$,
then it is evident
\begin{equation}
f = \int d \lambda \lambda \chi_{A_{\lambda}}.
\end{equation}

\subsubsection{The decoherence functional}
A {\em decoherence functional} $\Phi$ \footnote{ Note that we changed notation for the
decoherence functional from $d$, which is the standard in the bibliography to $\Phi$
and was employed in the previous sections. This was done for reasons of notational
convenience ($d$ tended to be confused with differentials). The letter  $\Phi$ stands
for phase.} is a map from ${\cal C} \times {\cal C} \rightarrow  {\bf C}$, such that
the following conditions are satisfied
\\ \\ \\
B1. {\em Null triviality:} For any $A \in {\cal C}$, $\Phi(\emptyset,A) = 0$. \\
 In terms of our interpretation of the off-diagonal elements of the decoherence functional
 as corresponding to Pancharatnam phases, there can be no phase measurement if one
 of the two beams that have to be interfered is absent. \\ \\
\\
B2. {\em Hermiticity:} For $A,B \in {\cal C}$, $\Phi(B,A) = \Phi^*(A,B)$. \\
Clearly the phase difference between two histories becomes opposite if we exchange the
 sequence, by which these histories are considered.
\\ \\ \\
B3. {\em Positivity:} For any $A \in {\cal C}$, $\Phi(A,A) \geq 0$. \\
This amounts to the fact that the diagonal elements of the decoherence functional
 are interpreted as probabilities (albeit non-additive). Operationally probabilities
  are defined by the number of times a particular event occurred in the ensemble and  by
   definition they can only be positive.
\\ \\ \\
B4. {\em Normalisation:} $\Phi(\Omega, \Omega) = 1$. \\
Clearly, if no measurement takes place the intensity of the beam would never change.
\\ \\ \\
B5. {\em Additivity:} If $A, B, C \in {\cal C}$ and $A \cap B = \emptyset$,  then
$\Phi(A \cup B,C) =
\Phi(A,C) + \Phi(B,C)$. \\
There is no intuitive operational reason, why this should be the case. This property
is equivalent to the superposition principle of quantum theory and we can consider
that it is forced upon us by experimental results. Of course, this is the property
that makes the decoherence functional the natural object to use.
\\ \\ \\
B6. {\em Boundedness:} For all $A,B \in {\cal C}$, $ |\Phi(A,B)| \leq 1 $. This arises
from the operational procedure for the determination of $|\Phi(A,B)|$ \cite{Ana02}.
\\
\\
These axioms are an adaptation of the axioms written by Isham and Linden \cite{IL94}
for the case of consistent histories.

There are two points one needs to make regarding these axioms. First, the properties
of $\Phi$ are identical to the matrix elements of a density matrix over some
continuous basis (if this is identified with $\Omega$). So {\em the difference of
quantum  from stochastic processes is that they have a density matrix rather than a
probability measure over the sample space of histories}.

The second point is that the properties of the decoherence functional are these of a
complex probability measure on $\Omega \times \Omega$. So we need not construct any
different mathematics for the development of the theory: standard measure theory will
suffice. In particular, we can use the Radon-Nikodym and Kolmogorov theorems that are
very important in  classical probability theory. Since $\Phi$ acts on events its
action can be extended to all measurable functions on $\Omega$, i.e. we can write it
as a map $\Phi : F(\Omega) \otimes F(\Omega) \rightarrow {\bf C}$.

Now, if on $\Omega$ there is a Lebesque measure $dx$, we can write $\Phi$ in terms of
s density $\upsilon$, i.e.
\begin{equation}
\Phi(dx,dx') = \upsilon(x,x') dx dx' .
\end{equation}

In the trivial case that $\Omega$ refers only to a moment of time, it is easy to see
that $\upsilon$ is written in terms of the density matrix as
\begin{equation}
\upsilon(x,x') = \rho(x,x') \delta(x,x')
\end{equation}
in a continuous basis (like position) or
\begin{equation}
\upsilon(x,x') = \langle z|\rho|z' \rangle \langle z'|z \rangle,
\end{equation}
in an overcomplete basis like the coherent states.

\subsection{Quantum processes}

We are interested in the non-trivial case, whenever $\Omega$ is a space of histories,
i.e. it is a suitable chosen subset of $\times_t \Gamma_t$. We shall assume $t$ to
take values in some interval of the real line, or the real line itself. An element of
$\Omega$ will then be a path on $\Gamma$ and will be written written as $z(\cdot)$.

Given a function $f$ on $\Gamma$, we can define a family of functions $F_t$ on $\Omega
$ indexed by $t$ as
\begin{equation}
F_t[z(\cdot)] = f(z(t)).
\end{equation}

In analogy with the definition of a stochastic process, a {\em quantum process} is
defined as a triple $(\Omega, \Phi, F^a_t)$, where
\\
- $\Omega$ is a path space \\
- $\Phi$ is a decoherence functional satisfying the relevant axioms \\
- $F^a_t$ is a selected family of observables indexed by $t$ (not necessarily of the type (4.7)). \\ \\
\\
From the specification of the family of observables $F^a_t$ we can define the mixed
correlation functions $G^{n,m}$ as
\begin{eqnarray}
G^{n,m}(a_1,t_1; a_2,t_2; \ldots ; a_n,t_n| b_1,t'_1; b_2,t'_2; \ldots ; b_m,t'_m) =
\\ \nonumber
 \Phi(F^{a_1}_{t_1} F^{a_2}_{t_2}
\ldots F^{a_n}_{t_n}, F^{b_1}_{t'_1} F^{b_2}_{t'_2} \ldots F^{b_m}_{t'_m}) .
\end{eqnarray}
Here $G^{n,0}$ are the time-ordered correlation functions, $G^{0,m}$ are the
anti-time-ordered and $G^{n,m}$ ones containing mixed entries.

From the hierarchy  $G^{n,m}$ of correlation functions associated to $F^a_t$ we can
define the corresponding
 generating functional $Z_F[J_+,J_-]$, which is
written in terms of the sources $J^{a}_+(t), J^{a}_-(t)$ as
\begin{eqnarray}
Z_F[J_+,J_-] = \sum_{n=0}^{\infty} \sum_{m=0}^{\infty} \frac{i^n (-i)^m}{n! m!}
\hspace{5cm}
 \nonumber \\ \times \sum_{a_1,\ldots a_n} \sum_{b_1,
\ldots, b_m}  \int dt_1 \ldots dt_n dt'_1 \ldots dt'_m \hspace{3.5cm}
\nonumber \\
\times G^{n,m}(a_1,t_1; \ldots ; a_n,t_n| b_1, t'_1; \ldots ; b_m, t'_m) \hspace{2cm}
 \nonumber \\
\times J^{a_1}_+(t_1) \ldots J^{a_n}_+(t_n) J^{b_1}_-(t'_1) \ldots
 J^{b_m}_-(t'_m)
\end{eqnarray}

This is known as the closed-time-path (CTP) generating functional. It was first
introduced by Schwinger \cite{Schw61} and Keldysh \cite{Kel64}. It is particularly
relevant to the discussion of systems, in which there does not exist the symmetry of
time translation (open quantum systems, field theory in curved spacetime etc). Clearly
the CTP generating functional can be written as
\begin{equation}
Z_F[J_+,J_-] = \Phi(e^{i F \cdot J_+}, e^{-i F \cdot J_-}),
\end{equation}
where $F \cdot J_{\pm} : = \int dt \sum_i F^{i}_t J_{\pm}Бо(t)$.

\paragraph{Defining $\Phi$}
$\Omega$ is in general a path space. In order to define $\Phi$ on it, we treat it as a
complex measure and employ an analogue of {\em Kolmogorov 's theorem} for probability
measures on path spaces. Namely, that $\Phi$ is uniquely determined by the
discrete-time versions of the decoherence functional: i.e. by  a hierarchy of
distribution functions
\begin{equation}
\upsilon^{n,m}(z_1,t_1; z_2,t_2; \ldots ; z_n,t_n|z_1', t_1'; z'_m,t'_m)
\end{equation}

The functions of this hierarchy need to satisfy the  {\em (Kolmogorov) additivity
condition:}
\begin{eqnarray}
\int dz_{n+1} \upsilon^{n+1,m}((z_1,t_1; z_2, t_2; \ldots; z_n,t_n; z_{n+1},t_{n+1}|
z'_1, t'_1; z'_2,
t'_2; \ldots ; z'_m, t'_m) = \nonumber \\
 \upsilon^{n,m}(z_1,t_1; z_2, t_2; \ldots; z_n,t_n| z'_1, t'_1;
z'_2, t'_2; \ldots ; z'_m, t'_m). \hspace{2cm}
\end{eqnarray}

The point is that a measure (hence a decoherence functional) for the continuous-time
case can be determined by the specification of only discrete-time expressions.

\subsection{The kinematical process}

One can construct quantum processes starting from standard quantum theory, by
employing the coherent states. In general, the coherent states provide a  from a
symplectic manifold $\Gamma$ to the projective Hilbert space: $z \in \Gamma
\rightarrow |z \rangle \langle z|$. If $f$ is a function on $\Gamma$ then one can
define a corresponding operator on the Hilbert space as
\begin{equation}
A = \int dz f(z) |z \rangle \langle z| .
\end{equation}
This is not the only choice, but as we shall see it is the most natural.

We want first to define the {\em kinematical process}, i.e. the quantum process that
corresponds to a system with vanishing Hamiltonian.

Having $\Gamma$ one can define the space $\Omega$ of continuous paths on $\Gamma$ and
the family of functions $F^a_t$ associated to the $f^a$ of equation (4.13). All that
is missing from the definition of  a quantum processes is the specification of a
decoherence functional. This is achieved by specifying the hierarchy of ordered
distribution functions $\upsilon^{n,m}$. To do so, we write the time instants in terms
of their ordering $t_1 \leq t_2 \leq \ldots t_n$, and $t'_1 \leq t'_2 \leq \ldots \leq
t'_m$. If we  write $\hat{\alpha}_z = | z \rangle \langle z |$ we will have
\begin{eqnarray}
\upsilon^{n,m}_{z_0}(z_1,t_1; z_2,t_2; \ldots ; z_n, t_n|z'_1,t'_1;
z'_2, t'_2; \ldots ; z'_m,t'_m) = \nonumber \\
Tr    \left( \hat{\alpha}_{z_n} \hat{\alpha}_{z_{n-1}} \ldots \hat{\alpha}_{z_2}
\hat{\alpha}_{z_1} \hat{\alpha}_{z_0} \hat{\alpha}_{z'_1} \ldots
\hat{\alpha}_{z'_{m-1}}
\hat{\alpha}_{z'_m} \right) = \nonumber \\
\langle z'_m| z_n \rangle \langle z_n| z_{n-1} \rangle \ldots \langle z_2| z_1 \rangle
\langle z_1| z_0 \rangle \langle z_0| z'_1 \rangle \langle z'_1| z'_2 \rangle \ldots
\langle z'_{m-1}| z'_m \rangle .
\end{eqnarray}
Let us now note the following concerning the kinematical process.
\\ \\
1. The expression for the distribution function factorises in products of the form
$\langle z| z' \rangle$. The knowledge of this inner product, suffices to fully
determine the kinematical process. In fact, the distribution function $\upsilon^{n,m}$
is known as the $n+m+1$ Bargmann invariant \cite{MuSi93}.
\\ \\
 2. The distributions $\upsilon^{n,m}$ do not depend on the
values of time $t$, only on their ordering. The same is true for $t'$. More than that,
if we consider the following cyclic ordering for the time instants $t_0 \rightarrow
t_1 \rightarrow t_2 \rightarrow \ldots \rightarrow t_n \rightarrow t'_m \rightarrow
\ldots \rightarrow t'_2 \rightarrow t'_1 \rightarrow t_0$, the distributions are
invariant if we consider any time as origin and then proceed cyclically along the
arrows. In other words, the kinematic process manifests the symmetry of a {\em closed
time path}.
\\ \\
3. Unlike stochastic processes, in which the kinematical process is trivial (the
hierarchy of distribution functions consists only of products of delta functions),
quantum processes manifest all quantum mechanical behaviour (interferences etc)
already at the kinematical level. The introduction of dynamics requires only the
minor modification of substituting  $\langle z|e^{-iH(t-t')}|z \rangle$ for $\langle
z| z' \rangle$ in equation (4.14).
\\ \\
4. Let us consider that the process being defined in the time interval $[0,\tau]$ and
consider the  distribution function $\upsilon^{n,m}$ for large values of $n$ and $m$.
Take for simplicity $n = m = N$. Choose also the time  instants  such that $|t_i -
t_{i-1}| \leq \delta t = \tau/N$ for all $i$ and similarly for $t'$. Also, let $t_n =
t'_m = \tau$. Then we have a discretised approximation to a decoherence functional for
continuous paths $z(\cdot), z'(\cdot)$, which  for $N \rightarrow \infty$ would
converge to
\begin{equation}
\Phi(z(\cdot),z'(\cdot)) = e^{- i \int_{C}  \langle z|d|z \rangle } + O(\delta t^2) =
e^{i \int_C A} + O(\delta t^2),
\end{equation}
where $C$ is the closed path obtained by appending the path $z'(\cdot)$ with reverse
orientation at the end of $z(\cdot)$. The distribution function for the decoherence
functional then converges at the large $N$ limit to the holonomy of a $U(1) $
connection
 on $\Gamma$, the same geometrical object that is introduced in the geometric quantisation scheme.  Of course, this convergence
is to be interpreted with a grain of salt as the  support of the decoherence
functional is primarily not on differentiable paths, for which the holonomy is
rigorously defined.

\subsection{Quantum differential equations}

In any theory that is based on histories, one can sharply distinguish between two
different aspects of temporal symmetries. This has been pointed out by Savvidou
\cite{Sav99a} and forms one of the basic features of the histories quantisation
programme. We can define a purely kinematical time translation, by which the time
translation is effected as
\begin{equation}
F_t \rightarrow F_{t+s},
\end{equation}
and there exists the dynamical time-translation generated by the Hamiltonian. These
are completely distinct and correspond to the different functions of the notion of
time in the physical theory (causal ordering/kinematics  vs change/ dynamics).

As we showed earlier, the characteristic quantum mechanical behaviour exists already
at the kinematical level and the decoherence functional depend only on the causal
ordering of the events constituting the histories. We are, therefore, tempted to take
the kinematical  process as basic and seek to write a general quantum process in terms
of it. In effect, we want to view the kinematical  process as the analogue of the
Wiener process in classical probability theory: a general object through which any
other process can be defined.

In classical stochastic processes the relation between a process $x$ and the Wiener
process $W$ \footnote{ Denoting the Wiener process as $W$  is a shorthand: the Wiener
process refers to the underlying sample space, the particular probability measure and
the basic variables $W^a$.} is expressed in terms of {\em stochastic differential
equations}, known in physics as Langevin equations. They are, in general, of the form
\begin{equation}
dx^a + f^a(x) dt = dW^a(t).
\end{equation}

In effect the Wiener term acts as a random driving force on a deterministic equation.
{\em Can we do the same for the case of the quantum processes?}

It turns out that we can. Using explicitly the distinction between the kinematical and
dynamical time translations, we referred to earlier we can write an analogous {\em
quantum differential equation}:
\begin{equation}
dz^a + f^a(z) dt = d \xi^a(t),
\end{equation}
where $z^a$ are variables (coordinates on $\Gamma$)  undergoing a quantum process with
a given Hamiltonian $H$, $f^a$ are some particular functions on phase space depending
on the Hamiltonian \cite{Ana02} and $\xi^a$ are variables on the phase space (the same
functions as $z^a$) but corresponding to a {\em kinematical process}. As in the
classical case, the differentials have t be interpreted with care as in general the
sample paths $z^a_t$ are not differentiable. This equation is written at the same
level of rigor as the standard stochastic differential equations.

In line with Einstein's remark we gave in section one,  we want to remark on the
appealing possibility that equation (4.18) can be interpreted as referring to an
individual system in analogy to the classical Langevin equations. That is, we can
consider that equation (4.18) refers to an individual system (a particle), which is
found within a ``fluctuating environment'', that induces the ``random forces''
$\dot{\xi}^a(t) $. However, these forces are not distributed according to a classical
probability distribution, but according to the kinematic processes (and are possibly
geometrical in origin).

We are not  in a position to argue, whether this interpretation of equation (4.18)
should be taken seriously or not. The reasons are partly mathematical and partly
physical: from the mathematical side we need to verify that such equations  are more
than empty symbols: is it actually a type of equation that can admit solutions? We
hope to justify such equations by adopting the theory of stochastic integrals (of Ito)
in the quantum context. From a physical point of view, even though we are committed to
finding a description for the individual quantum system, the picture of a particle
moving under random forces is not necessarily our first choice: it is  too classical
and there is no geometric naturality (the functions $f$ in (4.18) have no apparent
geometric interpretation).

Nonetheless, equation (4.18) has large theoretical interest: it {\em demonstrates}
that it is possible in principle to unravel the statistical description of quantum
theory into a description of individual system. The description in terms of quantum
differential equations may not be fundamental, it probably is not the physically
correct way to approach individual system, but it {\em proves a point}: {\em a
description of individual quantum systems that fully agrees in the ensemble statistics
with standard quantum theory is not impossible}.

Moreover, we  would like to see, whether it would be possible to simulate its
solutions  numerically as we can do with stochastic processes. This would provide  a
way of generating actual trajectories for individual quantum systems.

\section{Reconstruction theorem}

In the previous section, we showed how to obtain quantum processes starting from
quantum theory. Now, we want to invert this procedure and ask how one can obtain
standard quantum theory starting from a generic quantum process, that satisfies the
axioms stated in section .

Our result is simpler than we expected. We essentially found that
 we can uniquely determine the quantum mechanical Hilbert space, the observables and the evolution equations from the ingredients of a quantum process, if this process satisfies the (analogue of) the {\em Markov property}. The Markov property roughly  states that if the state of the system  (i.e the restricted decoherence functional at a moment of time) is completely specified, then it contains sufficient information to determine the state of the system at any subsequent moment of time.

The Markov property implies (this is often  taken as its defining property) that the
distribution functions that define decoherence functional can be written as
\begin{eqnarray}
\upsilon^{N+1,N+1}(z_0,t_0;z_1,t_1; \ldots ; z_{N},t_{N-1}|z'_0,t_0; z_1,t_1;
 \ldots ; z'_{N-1},t_{N-1}) =
\nonumber \\
\upsilon(z_{N},z'_{N};t_{N}|z_{N-1},z'_{N-1};t_{N-1})
 \ldots \upsilon(z_1,z'_1;t_1|z_0,z'_0;t_0)
\rho_0(z_0,z'_0), \label{factorisation}
\end{eqnarray}
in terms of a propagator $\upsilon(z_1,z_2;t|z'_1,z'_2;t')$ and an initial ``state''
at $t = 0$ \footnote{ Note that we have written only the diagonal elements of the
hierarchy of functions. But it can be easily shown that they can be used to construct
the full hierarchy $\upsilon^{n,m}$ by virtue of the Kolmogorov additivity
condition.}.

The propagator $\upsilon$ needs to satisfy the {\em quantum Chapman-Kolmogorov
equation}:
\begin{equation}
\upsilon(z_1,z_1';t|z_0,z'_0;s) = \int dz dz' \upsilon(z_1,z_1';t|z,z';s')
\upsilon(z,z';s'|z_0,z'_0; s).
\end{equation}

Now, what we have proven is a {\em reconstruction theorem}, which can loosely be
stated as follows.
\\ \\
{\bf Reconstruction theorem:} Assume we have a stochastic process $(\Omega, \Phi, F^a_t)$ that satisfies the Markov property. If in addition \\
i. the propagator is a smooth function of its arguments and the time entries, \\
ii. the process is time-homogeneous, \\
iii. the process is time-reversible ,\\
then we can reconstruct the quantum mechanical Hilbert space and the Heisenberg evolution equations. \\ \\
{\bf Sketch of the proof:} Time homogeneity means that the propagator depends only on
the time difference $t-t'$, hence can be written as $\upsilon_t(z_1,z_2|z'_1,z'_2)$.
Time-reversibility is defined as $\upsilon_t^*(z_1,z_2|z'_1,z'_2) =
\upsilon_{-t}(z_1,z_2|z'_1,z'_2)$. It is easy to show, that this implies that
$\upsilon $ is factorised as $\upsilon_t(z_1,z_2|z'_1,z'_2) = \psi_t(z_1|z'_1)
\psi^*_t(z_2|z'_2)$, in terms of another kernel that also satisfies a version of the
Chapman-Kolmogorov equality
\begin{equation}
\psi_t(z|z') = \int dz'' \psi_{t-s}(z|z'') \psi_s (z''|z').
\end{equation}
The condition (i) is important. It ensures that when $t \rightarrow 0$, $\psi$ remains
a nice (i.e.  differentiable) function (not a distribution), say $ \chi(z|z') :=
\psi_0(z|z')$, hence {\em the kinematical process will not be trivial}. In this case
the Chapman-Kolmogorov identity states that
\begin{equation}
\chi(z|z') = \int dz'' \chi(z|z'') \chi(z''|z'),
\end{equation}
hence $\chi$ defines a projection operator $E$ on $ {\cal L}^2(\Gamma)$. The range of
$E$ is the quantum mechanical Hilbert space $H$. Moreover, the dynamics encoded in
$\psi_t$ correspond to an one-parameter group of unitary transformations that commutes
with $E$ and can thus be projected on $H$ giving rise to Hamiltonian evolution.

One can also use a standard $GNS$ construction \cite{KlSk85} to construct a family of
coherent states $|z \rangle$ on $H$ such that
\begin{equation}
\langle z|z' \rangle = \chi(z|z').
\end{equation}

In this case a function $f$ on $\Gamma$ is mapped into an operator $A = \int dz |z
\rangle \langle z |$. In particular, a phase space cell $C$ is mapped to a positive
operator $P_C = \int_C dz |z \rangle \langle z|$.
\\ \\ \\
One remark needs to be made at this point. Our procedure so far is axiomatic and not
constructive. If one wants to explicitly construct the quantum process one needs to
write the coherent states propagator. This can be rigorously defined by a phase space
path integral, in which the Planck's constant enters through a Riemannian metric on
the phase space that is employed for regularisation \cite{Kla88}.

\section{Interpretational issues}

The difference between quantum processes  and standard quantum theory lies only in the
determination of which object correspond to sharp events. Quantum mechanics admits
projection operators, while the theory of quantum processes admits phase space cells.
These are represented by a positive operator-valued-measure $C \rightarrow \hat{C} =
\int_C dz |z \rangle \langle z|$, for any measurable subset $C$ of $\Gamma$.

The question then arises, which of the basic principles of quantum theory is (are )
violated by this change and whether this violation has empirical consequences.

It can be shown \cite{Ana02} that the  only difference between the theory of quantum
processes and standard quantum mechanics  is (what we shall call) the {\em spectral
principle}
\\ \\
{\em The possible values for an observable correspond  to the points of the spectrum
of the corresponding operator.}
\\ \\
It is  a corollary of this postulate, that a proposition about possible values of an
observable is represented by a projection operator.

 Now, in a quantum process the
spectrum of an operator is simply not relevant to the values of the corresponding
observable, because at the fundamental level observables are functions on the history
space $\Omega$. Clearly there is little difference as far as observables with
continuous spectrum are concerned (position, momentum etc). The difference lies, of
course, in the case of observables with discrete spectrum.

The case of discrete spectrum is, in fact, what has given quantum phenomena their
name, as it is this through the discrete spectrum of operators that the paradigmatic
quantum behaviour is manifested: historically it was the black body radiation, the
photoelectric effect  and the Bohr's atom transitions that put discreteness as a basic
feature of the new mechanics. For this reason the spectral postulate was highlighted
in all early work of quantum theory: it provided  a simple  solution to the problems
that had faced a generation of physicists. Later mathematical development -namely the
spectral theorem - offered this postulate the additional justification of mathematical
elegance.

It would seem that this is one of the most solid postulates of quantum  mechanics, the
last one to be taken away from any possible modification of the theory. After all it
provides the solution to the physical puzzles that led to quantum  mechanics. However,
as we are going to argue it is the postulate of quantum theory that is {\em least}
justified empirically, when taken by itself.

To see this we shall consider the case of atom spectroscopy, which has been
 historically the main arena justifying the spectral postulate.
 When we study  the
electromagnetic radiation emitted from atoms, we see that the intensity of the
electromagnetic field has peaks in particular discrete values of the frequency. Then
assuming energy conservation, the photons are viewed as arising from  a transition
between two "states" of an atom , each of which is characterised by a sharp value of
energy. The fact that we measure  a  number of sharp peaks rather than a smoother
distribution of field intensity plotted versus frequency, leads to the conclusion that
the possible values of atom's energy are discrete. If we take that this experiment
measures the energy of the atom, then we have discrete values for the energy,
something that is naturally explained in terms of the spectral postulate: in any
individual measurement only points of the spectrum of the operator are obtained.

We believe that this is a fair summary of the argument that leads to the acceptance of
the spectral postulate  in this particular context. We shall now see, that the
conclusions of the argument is by no means necessary. Let us first make the too
obvious remark, that the measurement of the intensity peaks never yields sharp values,
rather only peaks with finite width. The width is due not only to experimental errors,
but comes fundamentally from the time-energy uncertainty relation. Hence, it is only
in an idealisation that the atom's energy values are discrete.

However, the most important argument is that the description in terms of atom
transitions {\em is semiclassical rather than quantum}. What we measure in
spectroscopy is  the energy/frequency of the electromagnetic field. We typically
assumed that the emitted photons are incoherent (both in the classical and the quantum
sense), so that the emitted electromagnetic field can be considered as an ensemble of
photons. Then, we can idealise the experiments as setting filters that allow only very
narrow frequency (energy) range to pass and measure the intensities. The whole
experiment is then fully described by energy measurements of the photons. One can give
an equivalent description in terms of the electromagnetic fields. So the actual
observables that correspond to the set-up of the experiment is photon energies or
fluxes, {\em not atomic energies}. And these energies can be described by continuous
variables in either quantum theory or in the quantum process description.

The attribution of discrete energy values to the atom comes from a semiclassical {\em
picture} of the atom/field interaction; it involves a mixture of old quantum theory
concepts (orbitals, transitions), with the framework of mature quantum theory. This
picture is helpful for calculations, it provides an intuitive picture of the
interaction, but it is not fundamentally quantum mechanical. A precise treatment ought
to consider the combined system field-atom, interacting perhaps through QED and then
consider energy measurements of the electromagnetic field at particular spatial
locations. In such a description all information about the process (including the
atom's eigenvalues) would be found in the correlation functions of the electromagnetic
field: {\em but these are predicted by quantum processes in full agreement with
standard quantum theory}.

What we imply by this argument, is that historically the discrete values of
observables actually refer to the spectrum of the Hamiltonian, rather than any
arbitrary observable. The information about its eigenvalues is fully contained in the
correlation functions: once these are provided, we can read off any discretised
behaviour. In other words, {\em the discrete behaviour in quantum theory is not
fundamental or ontological, but arises due to particular forms of the dynamics.} This
is true even for spin systems: the "discrete" spin values are always measured in
conjunction with its coupling to some magnetic field.

\section{Conclusions}

Let conclude in the form of a summary:
\\ \\
1. We argued that complex numbers (or a U(1) invariance of probabilities) is an inherent and irreducible component of quantum probability. Their effect is the existence of statistical quantities that can be determined by experiment that have no analogue in quantum probability: these are the geometric phases that can be determined by comparing {\em two } distinct histories of the system. Compared to classical probability these phases correspond to novel operational concepts. \\ \\
2. The  consistent histories approach provides the best formalism to take the phase
information into account: there exists  an 1-1 map between observable quantities
(including phases) and mathematical objects. This comes from the relation of the
values of the decoherence functional to the Pancharatnam phase. We do not need,
however, to subscribe to the standard interpretation of consistent histories.
Throughout this paper we prefer to keep an operational perspective.
\\ \\
3. Taking phases as primitive objects of the formalism necessitates the use of
non-additive probabilities. Theories with non-additive probabilities can be described
by commutative observables (or classical logic or hidden variables) without violating
Bell's theorem or Kochen-Specker's theorem.
\\ \\
4. We can write a theory of quantum processes in analogy with the theory of stochastic
processes. This theory has a well-defined sample space \footnote{ In a recent paper,
Kent makes the very accurate remark that the quantum measurement problem should be
more accurately phrased as what is exactly the sample space of the quantum system
\cite{Kent02}. Our answer could have come out of  Bohr: the sample space is identical
to the one of the corresponding classical system.}
 (we choose the classical phase space). The only difference is that in the quantum case the relevant object is a ``density matrix'' in the space of paths. In this picture, the Schr\"odinger equation is an exact analogue of the Fokker-Planck equation. Moreover, we can unravel the statistical description to write (at least formally) quantum  equations. They can be thought of, as corresponding to individual systems; whether they correspond to real physics is doubtful, but they prove that Einstein's suggestion that {\em quantum theory is a statistical theory arising out of yet unknown physics for the individual system} is possible and not in conflict with any predictions of quantum theory.
\\ \\

5. Its physical implications aside,  it would be very interesting to see if the
solutions to quantum differential equations can be found numerically. We would be,
then, able to simulate the evolution of individual quantum systems.
\\ \\
6. Starting from quantum processes we get standard quantum theory, by assuming the
Markov property, time homogeneity, time-reversibility and the non-triviality of the
kinematical process. Hence, the structure of the Hilbert space necessitates the Markov
condition, which presupposes a background causal structure. In absence of this {\em
the Hilbert space is not necessary or even natural}. This could be the case in {\em
quantum gravity}.
\\ \\
7. Finally, we want to identify, how the spacetime symmetries are implemented in the
theory of stochastic processes: the relation with the histories quantisation programme
guided our choice of sample space. After all, our motivation is to find a covariant
description of quantum systems, that would allow us to tackle the quantisation of
gravity. What we would like to see is that the complex phases of quantum theory, would
be deeply intertwined with the spacetime structure (perhaps in a fashion analogous to
an old conjecture by Penrose?).

\section*{Acknowledgments}
I would like to thank the organisers of the conference and Edgard Gunzig, in
particular, for their invitation and the very gracious hospitality.

I would also like to thank N. Savvidou for long discussions and a fruitful interaction
in many aspects of this project.

The research was supported by a Marie Curie Fellowship of the European Commission. The
Commission is not responsible for any views expressed here.

\end{document}